\documentclass[a4paper]{article}
\usepackage{Odyssey2022}
\usepackage{epsfig,amssymb,amsmath}
\usepackage{url,booktabs}
\usepackage{cite}
\ninept

\title{Pretraining Approaches for Spoken Language Recognition: TalTech Submission to the OLR 2021 Challenge}

\name{Tanel Alumäe, Kunnar Kukk}

\address{Department of Software Science  \\
Tallinn University of Technology, Estonia \\
{\small \tt tanel.alumae@taltech.ee} }
\begin{document}
\maketitle
\begin{abstract}
This paper investigates different pretraining approaches to spoken language identification. The paper is based on our submission to the Oriental Language Recognition 2021 Challenge. We participated in two tracks of the challenge: constrained and unconstrained language recognition. For the constrained track, we first trained a  Conformer-based encoder-decoder model for multilingual automatic speech recognition (ASR), using the provided training data that had transcripts available. The shared encoder of the multilingual ASR model was then finetuned for the language identification task. For the unconstrained task, we relied on both externally available pretrained models as well as external data: the multilingual XLSR-53 wav2vec2.0 model was finetuned on the VoxLingua107 corpus for the language recognition task, and finally finetuned on the provided  target language training data, augmented with CommonVoice data. Our primary metric $C_{\rm avg}$ values  on the Test set are 0.0079 for the constrained task and 0.0119 for the unconstrained task which resulted in the second place in both rankings. In post-evaluation experiments, we study the amount of target language data needed for training an accurate backend model, the importance of multilingual pretraining data, and compare different models as finetuning starting points.
\end{abstract}

\section{Introduction}

Spoken language identification (LID) is the task of automatically identifying the language of an utterance. Speech-based LID is used as a pre-processing step in several applications, such as automatic call routing, multilingual spoken translation and human-machine communication systems, multilingual speech transcription systems and spoken document retrieval. SLR is also often used in the area of intelligence and security.

In order to encourage the research on multilingual phenomena and advance the development of language and dialect
recognition technologies, the Oriental Language Recognition (OLR) Challenge has been organized annually since 2016~\cite{olr2016,olr2017,olr2018,olr2019,olr2020}. The sixth OLR challenge, denoted by OLR 2021 Challenge \cite{wang2021olr}, included two LID tasks and two ASR tasks, both evaluating either constrained or unconstrained training scenarios.  The constrained  LID task is a cross-domain closed-set identification task  with 13 target languages. Only the data provided by the organizers can be used to build the system. The unconstrained LID task is a closed-set identification task with 17 languages. Here,  utterances obtained from real-life environments are used for evaluation.  Any data and models are allowed for system training and development. As in NIST LRE15, the OLR 2021 Challenge uses $C_{\rm avg}$ as the principle evaluation metric.

The Tallinn University of Technology (TalTech) team participated in both language identification tasks. We relied on transfer learning: in the constrained task, we first trained a multilingual automatic speech recognition (ASR) model and finetuned it for language recognition, similarly as proposed in \cite{wang21z_interspeech}. In the unconstrained task, we finetuned the multilingual XLSR-53 wav2vec2.0 model \cite{conneau2020unsupervised} first on the VoxLingua107 dataset \cite{valk2021slt} and then on the target language/dialect data.

\section{Previous work}

Various pretraining and transfer learning schemes have been a popular approach to LID for a long time. In the phonotactic approach, one first trains a multilingual \cite{matejka2005phonotactic} or several monolingual phone recognizers \cite{zissman1996comparison} on transcribed training data. Streams of phone symbols are then recognized from test utterances. Phone stream from a test utterance is finally classified using language-specific phone n-gram models.

Another successful pretraining strategy is based on bottleneck features (BNFs) \cite{matejka2014neural,jiang2014deep,fer2015multilingual}.  Transcribed monolingual or multilingual spoken data is used for training a DNN-based acoustic model that has a narrow bottleneck layer close to the output layer. After training, the layers after the bottleneck layer are discarded and the output from the bottleneck layer is used as a frame-level feature extractor. Bottleneck features provide impressive performance improvements with both i-vector as well as neural network based frontends \cite{snyder2018spoken}.  The best-scoring submission to the OLR 2020 Challenge tasks 1 and 3 was also based on bottleneck features, extracted from an end-to-end multilingual Conformer-based ASR model \cite{duroselle2021language}.

Several works have shown that finetuning a pretrained feature extractor with the rest of the backend model for spoken language identification gives additional benefits with regard to using a frozen model for extracting bottleneck features. In \cite{wang2021end}, the shared encoder of a multilingual Conformer-based ASR model was finetuned for dialect identification, achieving strong results in the second task of the OLR 2020 Challenge. Recently, self-supervised models trained on large amounts of untranscribed data have been reported to achieve large improvements on language recognition. In \cite{tjandra2021improved} it was shown that multilingual wav2vec2.0 model XLSR-53 can be finetuned for language recognition using very little training data. XLS-R \cite{babu2021xlsr}, an improved version of the XLSR-53 wav2vec2.0 model, achieved state-of-the-art results on the VoxLingua107 development set.

\section{Task 1: Constrained Language Identification}

Task 1 of the OLR 2021 Challenge is a closed-set identification task with 13 target languages.  Training and test data is recorded in various different  environments, making it a cross-domain identification task. Only speech data provided by the organizers is allowed to be used for training. Non-speech data from external sources is allowed for data augmentation purposes.

\subsection{Data}

\begin{table}[tb]
\caption{Task 1 training and development data.}
\label{tab:task1-data}
\begin{tabular}{l|rr|rr|r}
\toprule
         & \multicolumn{2}{l|}{All training data} & \multicolumn{2}{l|}{Transcr. data} & Dev \\
\midrule         
Language         & \#Utts             & \#h            & \#Utts            & \#h          & \#Utts        \\
\midrule         
Kazakh    & 10748            & 18.2           & 4192            & 9.5           & 0           \\
Hokkien   & 9305             & 16.5           & 5163            & 15.6          & 1998        \\
Shanghainese & 9472             & 12.8           & 5374            & 12.1          & 1796        \\
Sichuanese  & 9487             & 11.5           & 5378            & 10.9          & 1798        \\
Tibetan    & 19480            & 18.0           & 5378            & 11.9          & 0           \\
Uyghur    & 13497            & 25.0           & 5382            & 13.6          & 0           \\
Cantonese    & 18572            & 21.5           & 5705            & 7.6           & 4353        \\
Indonesian    & 16053            & 18.9           & 5733            & 7.5           & 1800        \\
Japanese    & 17728            & 15.4           & 5734            & 5.8           & 4116        \\
Korean    & 17597            & 16.8           & 8800            & 6.0           & 3725        \\
Russian    & 14581            & 20.6           & 8967            & 9.9           & 3744        \\
Vietnamese    & 16902            & 21.0           & 8982            & 8.4           & 1800        \\
Mandarin    & 19302            & 22.0           & 11094           & 7.6           & 3600       \\
\bottomrule
\end{tabular}
\end{table}

We used all the provided training data for the 13 target languages for training the models, except for the OLR 2020 test data, that was used as a development set. A subset of the training data comes with transcripts. This data was used for pretraining the ASR model. The amount of data for each target language is summarized in Table~\ref{tab:task1-data}.
The background noises in the Freesound portion of the MUSAN corpus \cite{musan2015} and  simulated small, medium and large room impulse responses \cite{ko2017study} were used for data augmentation.

\subsection{Models}

In the constrained task, our system was a combination of four models. One of the models was a Resnet-style model, trained on the provided training from scratch. The other three were finetuned from Conformer-based multilingual ASR models, with different pooling methods and training criteria.

\subsubsection{Resnet-style model}

\begin{figure}[tbh]
\centering
\includegraphics[width=0.6\linewidth]{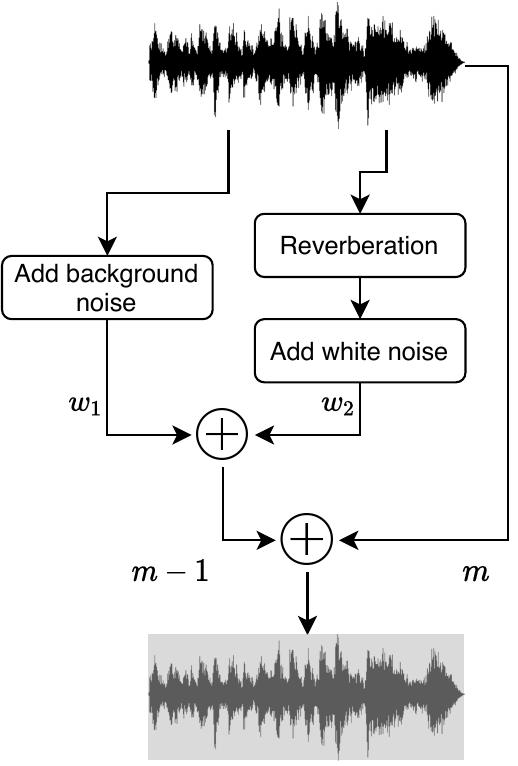}
\caption{{\it Sample augmentation pipeline in AugMix (largely based on \cite{hendrycks2019augmix}). The number and type of transforms in each augmentation path is randomly sampled. The augmented signals are mixed using randomly sampled $w_1$ and $w_2$. The mixed augmented signal is finally interpolated with the original signal, using a sampled weight $m$.}}
\label{fig:augmix1}
\end{figure}

The Resnet-style model is derived from the x-vector paradigm~\cite{snyder2018x,snyder2018spoken}, with several enhancements. During training, we apply on-the fly data augmentation using AugMix \cite{hendrycks2019augmix}, by randomly distorting the training data using a mix of reverberation and noise augmentation. A clean training segment is cloned into several copies. A different randomly drawn series of augmentation transforms, each possibly with random parameters, is applied serially to each of the copies. Then, the augmented copies are mixed with each other (using randomly sampled weights) and the resulting super-augmentated sample is finally mixed with the original clean sample, using a randomly sampled interpolation coefficient (see Figure \ref{fig:augmix1}).  The benefit of this method, compared to using pre-generated static augmentations, is that there is a lot of variety in the training data: each training sample is a result of several random transforms, applied in random order and with a random weight. 

For frame-level feature extraction, we use the Resnet34 \cite{cai2018exploring, he2016deep} architecture where the basic convolutional blocks with residual connections are replaced with squeeze-and-excitation modules \cite{hu2018squeeze,zhou2019deep}. 
The statistics pooling layer that maps frame-level features to segment level features is replaced in our model with a multi-head attention layer \cite{bahdanau2014neural} that has been shown to provide superior performance  \cite{zhu2018self, rahman2018attention, okabe2018attentive, safari2019self}. From among many variants of multi-head attention used in previous studies, we employ the one described in \cite{zhu2018self}: frame level representations are first mapped to $N_{\rm att}$ outputs ($N_{\rm att}=128$ in our model), using a $1\times1$ convolution and a ReLU nonlinearity; from this representation, each attention head (we used  $N_{\rm heads}=5$ heads) computes it's own softmax-based weight distribution over the input utterance; finally, weighted mean and standard deviation are computed over the frame level features for each head, resulting in $N_{\rm heads} \times 512 \times 2$ segment-level representations.

The structure of the embedding model is summarized in Table \ref{tab:architecture}. The variables $F$ and $T$ refer to the number of filterbanks and the number of time frames in the utterance. In all experiments, we used $F=30$.

\begin{table}[tb]
\caption{The neural network architecture of the Resnet model. Notation: \textit{SE/res} -- squeeze-and-attention block and residual connections; \textit{F} -- number of filterbanks in the input features; \textit{T} -- number of frames of input; \textit{FC} -- fully connected layer. }
\label{tab:architecture}
\footnotesize
\begin{tabular}{lllc}
\toprule
\textit{Layer}          &\textit{Spatial Size}        & \textit{\#Channels} &   \textit{Kernel}       \\
\midrule
Input          & $F \times T$            & 1              & -            \\
\multicolumn{4}{l}{\it{Frame level representations}} \\
Pre-resnet     & $F \times T$       & 64             & 7 $\times$ 7 \\
Res-block 1  & $F/2 \times T/2$   & 64       & $3 \times \begin{bmatrix} 3 \times 3 \\ 3 \times 3 \\ SE/res \end{bmatrix}$  \\
Res-block 2 & $F/4 \times T/4$   & 128        & $4 \times \begin{bmatrix} 3 \times 3 \\ 3 \times 3 \\ SE/res \end{bmatrix}$ \\
Res-block 3 & $F/8 \times T/8$   & 256       & $6 \times \begin{bmatrix} 3 \times 3 \\ 3 \times 3 \\ SE/res \end{bmatrix}$ \\
Res-block 4 & $F/16 \times T/16$ & 512        & $3 \times \begin{bmatrix} 3 \times 3 \\ 3 \times 3 \\ SE/res \end{bmatrix}$ \\
Post-resnet & $1 \times T/16$    & 512            & $F/16  \times 1$ \\ 
\multicolumn{4}{l}{\it{Segment-level representations}} \\
Pooling   & 1    & $5 \times 512 \times 2$ &  Attent. stats \\
Embedding        & 1    & 512     &  Dense \\
FC        & 1    & 512         &   Dense \\
Output       & 1    & \#Languages         &   Softmax \\
\bottomrule
\end{tabular}
\end{table}

The models are implemented in PyTorch \cite{NEURIPS2019_9015} using a framework developed in our lab.

\subsubsection{Conformer ASR model}

The Conformer-based ASR model was trained on pooled provided training data that came with transcripts. As a development set, we used the OLR 2020 test data. We applied a number of text normalization steps to the transcripts before training: for Cantonese, Mandarin, Mandarin dialects and Japanese, all whitespace symbols were deleted. For all languages, all punctuation symbols were deleted. For Kazakh, the Arabic script was transliterated into Cyrillic.

The ASR model uses a byte-pair encoding (BPE) vocabulary of 20\,000 units, shared over all languages. This is much larger than a typical BPE vocabulary size for a monolingual model. However, since the BPE tokens are shared over 13 languages, it averages to around 1500 BPE token per language which is typical for low resource end-to-end ASR models. The multilingual ASR model is an encoder-decoder based model that uses Conformer as an encoder and Transformer as a decoder. Some important hyperparameters of the model are listen in Table \ref{tab:conformer}. The model was trained on speed-perturbed data and SpecAugment \cite{park2019specaugment} (without warping) was applied during training. The number of training epochs was 40. The model was trained on four GPUs, with a dynamic batch size of 30M frames per GPU, which translates to 4620 updates per epoch.
After every epoch, the model's performance on development data was measured, and the final model was averaged over 10 best-performing models.

\begin{table}[tb]
\caption{Hyperparameters of the Conformer ASR model}
\label{tab:conformer}
\centering
\begin{tabular}{lr}
\toprule
\textit{Conformer encoder} &  \\
\midrule 
Number of blocks & 12 \\
Linear dimensionality & 2048 \\
Dropout rate & 0.1 \\
Output size & 256 \\
Num. attention heads & 8 \\
\midrule
\textit{Transformer decoder} & \\
\midrule
Linear dimensionality & 2048 \\
Number of blocks & 6 \\
Num. attention heads & 4 \\
\midrule
\textit{Training} & \\
\midrule
CTC weight & 0.3 \\
Label smoothing & 0.1 \\
\bottomrule
\end{tabular}
\end{table}

After training of the ASR model, the encoder part of this model was taken as the backbone of the language recognition model. The encoder's outputs were fed through a pooling layer. We experimented with different pooling methods: attentive statistics pooling \cite{okabe2018attentive}, multi-head attention (MHA) pooling  \cite{india2019self}, global multi-head attention (GMHA) pooling \cite{wang2020multi}. Two fully connected layers with the ReLu non-linearity and BatchNorm were appended to the pooling layer. As a training criterion, we experimented with cross-entropy (CE) loss and additive angular margin (AAM) loss \cite{deng2019arcface}. The ASR encoder part of the model was trained together with the rest of the model, without any learning rate scaling. For some models, stochastic weight averaging (SWA) \cite{izmailov2018averaging} was applied during the last 30\% of training.

The language recognition model was trained on all available provided training data for the 13 target languages. Language embeddings were extracted from the first fully-connected layer after the pooling layer. The dimensionality of the embeddings was 512.

The Conformer-based model for ASR was trained using ESPnet \cite{watanabe2018espnet}. For pooling operations and backend scoring (see below), ASV Subtools \cite{tong2021asv} was used.

\subsubsection{Back-end modeling}

For backend scoring, we used a multinomial regression model. Language embeddings were normalized and mean-centered based on the training data. No LDA was used in this task. The logistic regression model was rebalanced using priors proportional to the inverse amount of data for each language, in order to remove any language bias from the model.

We combine the scores of various systems using calibrated combination weights. For finding the model combination weights, we optimize  the parameters of a linear model based on the log-likelihood cost metric (CLLR) on the development data, using L-BFGS as the optimizer. Our own Pytorch-based calibration implementation was used which is freely available\footnote{\url{https://github.com/alumae/sv_score_calibration}}.

\subsection{Results}

Our results for Task 1, together with a comparison to the official baseline and the winning team's results, are listed in Table~\ref{tab:task1-res}. Note that the Progress set is a 30\% subset of the Test set, hence the results should be very similar. However, some teams improved their system after submission to Progress set leaderboard was closed which explains the large difference.
It can be seen that transfer learning from an ASR model gives a massive boost to the system performance. Our best single model is based on the trained ASR Conformer encoder, used global multihead attention pooling, cross-entropy loss and stochastic weight averaging. Fusion of four models gives only a slight improvement both on development and progress sets. The fusion was also used in our official test set submission.

\begin{table*}[tb]
\centering
\caption{Results on Task 1 with various systems and their combination.}
\label{tab:task1-res}
\begin{tabular}{lll|ll|ll|ll}
\toprule
              &                 &           & \multicolumn{2}{c|}{\textit{Dev}} & \multicolumn{2}{c|}{\textit{Progress}} &   \multicolumn{2}{c}{\textit{Test}} \\
\textit{Backbone}      & \textit{Pooling}         & \textit{Criterion} & \textit{$C_{\rm avg}$} & \textit{EER}  & \textit{$C_{\rm avg}$}  & \textit{EER} & \textit{$C_{\rm avg}$}           & \textit{EER}          \\
\midrule
Resnet        & Attentive stats & CE        & 0.0515      & 6.90     & 0.0601        & 5.72        \\
ASR Conformer & MHA             & CE        & 0.0110       & 1.68     & 0.0101        & 1.15         \\
ASR Conformer & GMHA            & CE + SWA  & 0.0081      & 1.27     & 0.0080         & 0.91       \\
ASR Conformer & MHA             & AAM       & 0.0129      & 2.19     & 0.0133        & 1.68        \\
\midrule
\multicolumn{3}{l|}{\textbf{Fusion}}                  & 0.0078      & 1.20     & \textbf{0.0074}        &\textbf{0.86}  & 0.0079 & 0.86    \\
\midrule
Baseline \cite{wang2021olr} &  & &  & & 0.0826 & 9.04  & 0.0817 & 8.98  \\
Team X-Voice (ranked 1st) \cite{lyuant2021}  &  & &  & & \textbf{0.0062} &	\textbf{0.65} & 0.0025 & 0.27 \\
Team funspeech (ranked 3rd) \cite{wangfunspeech2021} & & &  & & 0.0080 & 0.91 & 0.0083 & 0.93 \\
\bottomrule
\end{tabular}
\end{table*}

\section{Task 2: Unconstrained Language Identification}

Task 2 of the challenge is also a closed-set language identification task. However, in this task the number of target languages is 17 and the test data is more challenging, originating from real-life environments. In this task, any data, including externally available pretrained models, are allowed to be used for system building.

\subsection{Data}

\begin{table}[tbh]
\centering
\caption{Task 2 training and development data. For languages marked with an asterisk, the data originates from Mozilla CommonVoice.}
\label{tab:task2-data}
\begin{tabular}{l|rr|r}
\toprule
         & \multicolumn{2}{l|}{Training data} & Dev  \\
\midrule         
Language & \#Utts             & \#Hours       & \#Utts        \\         
\midrule
Hokkien (Minnan)   & 9305             & 16.5           & 1998 \\
Shanghainese & 9472             & 12.8           & 1796 \\
Sichuanese  & 9487             & 11.5           & 1798 \\
Tibetan    & 19480            & 18.0           & 0    \\
Uyghur    & 13497            & 25.0           & 0    \\
Cantonese    & 18572            & 21.5           & 4353    \\
English$^\star$       & 15000            & 20.5           & 3000 \\
Hindi$^\star$       & 6812             & 8.1            & 2020 \\
Indonesian    & 16053            & 18.9           & 1800 \\
Japanese    & 17728            & 15.4           & 4116 \\
Korean    & 17597            & 16.8           & 3725 \\
Malay$^\star$       & 15000            & 26.2           & 1800 \\
Russian    & 14581            & 20.6           & 3744 \\
Telugu$^\star$       & 4004             & 5.1            & 444  \\
Thai$^\star$       & 15000            & 18.2           & 3000 \\
Vietnamese    & 16902            & 21.0           & 1800 \\
Mandarin    & 19302            & 22.0           & 3600 \\
\bottomrule
\end{tabular}
\end{table}

For training the backend and for final finetuning of the language embedding model, we constructed a custom training  set covering all 17 languages (see Table \ref{tab:task2-data}). For most languages, we used the data provided in the OLR 2021 Challenge training dataset. However, for English, Hindi, Malay, Telugu and Thai, there is no training data provided. For those languages, we used data from Mozilla CommonVoice \cite{ardila2019common}.  For languages with a lot of CommonVoice data, we limited the data to a random 15000 utterance subset.  The development set was compiled from OLR2020 test data and CommonVoice data for the languages not covered by the OLR data. We acknowledge that the development set does not reflect the characteristics of the test data very well, since it mostly contains read speech, while the test data originates from diverse real-life situations.

For training the language embedding models, we relied a lot on the recently released VoxLingua107 dataset. The VoxLingua107 dataset \cite{valk2021slt}\footnote{Available at \url{http://bark.phon.ioc.ee/voxlingua107/}} is a large-scale dataset for training spoken language identification models that work well on diverse real-life data.


VoxLingua107 was compiled from automatically scraped YouTube data. 
The data collection process consists of the following steps: first, semi-random trigram search phrases were generated from the Wikipedia text corpus of the particular language. The search phrases were used to retrieve YouTube videos whose title or description matched the search phrase. Text-based language  identification was used for filtering out the videos with the title and description likely not in the given language. Audio tracks of the videos were downsampled to 16\,kHz.
Speech activity detection and speaker diarization were applied for extracting segments from the videos that contain speech. Long speech segments were split into utterance-like subsegments of up to 20 seconds in length.
Data-driven post-filtering was used to remove segments from the database that were likely not in the given language, increasing the proportion of correctly labeled segments in the dataset to 98\%, based on crowd-sourced verification. Some numerical facts about the VoxLingua107 training data are given in Table \ref{tab:voxlingua107}.

Most of the Task 2 target languages are covered by VoxLingua107. The exceptions are the Mandarin dialects, Tibetan, Uyghur and Cantonese.

\begin{table}[tb]
\caption{Statistics about the VoxLingua107 dataset.}
\label{tab:voxlingua107}
\centering
\begin{tabular}{lr}
\toprule
Number of languages  & 107 \\
Total number of videos & 64110 \\
Total number of hours & 6682 \\
Average number of hours per language & 62 \\
Average number of utterances per language & 23709 \\
Total amount of audio (uncompressed, in GB) & 758 \\
\bottomrule
\end{tabular}
\end{table}

\subsection{Models}

\begin{table*}[tb]
\caption{Results on Task 2 with various systems.}
\label{tab:task2-res}
\centering
\begin{tabular}{l|ll|ll|ll}
\toprule
              & \multicolumn{2}{c|}{\textit{Dev}} & \multicolumn{2}{c|}{\textit{Progress}} & \multicolumn{2}{c}{\textit{Test}}\\
\midrule                                                         
\textit{Model}    & \textit{$C_{\rm avg}$}       & \textit{EER}        & \textit{$C_{\rm avg}$}        & \textit{EER}         & \textit{$C_{\rm avg}$}        & \textit{EER}    \\
\midrule                                                         
Resnet, trained on VoxLingua107                                  & 0.053      & 5.15      & 0.078         & 7.98        \\
Resnet, trained on VoxLingua107, finetuned on training data      & 0.015      & 1.53      & 0.055         & 5.27        \\
XLSR-53, finetuned on VoxLingua107                        & 0.017      & 1.76      & 0.016         & 1.69        \\
XLSR-53, finetuned on training data                       & 0.003      & 0.29      & 0.044         & 3.78        \\
\textbf{XLSR-53, finetuned on VoxLingua107, then on training data} & 0.006      & 0.78      & 0.012         & 0.93   & 0.012 & 0.94     \\
\midrule
Team X-Voice (ranked 1st)  \cite{lyuant2021}  & & & 0.017 & 1.85 & 0.004 & 0.42 \\
Team nisp\_speech (ranked 3rd) \cite{jinnisp2021}  & & &        &      & 0.032 & 3.23 \\
\bottomrule
\end{tabular}
\end{table*}

We experimented with using the XLSR-53 wav2vec2.0 model~\cite{conneau2020unsupervised} as the backbone of our language embedding model. XLSR-53 is a large pretrained model trained on unlabeled multilingual data. The model is trained by jointly solving a contrastive task over masked latent speech representations  and learning a quantization of the latents shared across languages.  The model contains a convolutional feature encoder that maps raw audio to latent speech representations which are fed to a Transformer network that outputs context representations. XLSR-53 is pretrained on  
56\,000 hours of speech data from 53 languages. 

We used XLSR-53 as follows: the outputs from the wav2vec2 model were fed through an attentive pooling  layer, a fully connected layer with ReLU and BatchNorm, and the final output layer, corresponding to the languages of the training set. During training, the learning rate corresponding to the XLSR-53 model was set to 0.01 times lower than the base learning rate. We experimented with three finetuning scenarios: using VoxLingua107, using the training data of the 17 languages (OLR + CommonVoice), and finetuning first on the VoxLingua107 data, followed by the training data for the 17 languages.

We also experimented with a Resnet model, trained on VoxLingua107, and then finetuned on the OLR training data.

The same on-the-fly data augmentation strategy as was used in Task 1 was also used here for all models.

\subsection{Back-end modeling}

As in Task 1, we used a multinomial regression model for backend scoring. Language embeddings were normalized and mean-centered based on the training data. The resulting embeddings were reduced to 50-dimensional vectors using LDA. No rebalancing of the model bias was used in this task, since it was found to slightly hurt the system performance on the progress set.

\subsection{Results}

The results of various systems are listed in Table \ref{tab:task2-res}. It can be seen that using XLSR-53 as the basis for finetuning models gives a big boost to the system performance, particularly on the  progress set. Our internally compiled development set had quite different performance trends, compared to the progress set. This is probably due to the fact that the development set contained clean dictated speech, whereas the progress set contained data ``from the wild''. 

The best performing-model on the progress set was the XLSR-53 model, finetuned first on VoxLingua107 and then on the training data that covered all 17 target languages and dialects. This system was also used in our official test set submission.

\section{Post-evaluation analysis}

In this section, we analyze various aspects of pretraining for spoken language identification.

Since at the time of writing this paper, the OLR 2021 Challenge test data for Task 2  was not available, we performed the experiments on Task 1 and Task 2 of OLR 2020. Note that rules of the OLR 2020 Challenge restrict the participants to use only the provided training data that includes data from the previous OLR challenges and the THCHS30 Chinese speech corpus \cite{wang2015thchs} which thereby also rules out using transfer learning. Therefore, we don't claim any state-of-the art results on OLR 2020 test data as we deviate from the official rules. 

Task 1 of the OLR  2020 Challenge is a closed-set identification task with six target languages (Cantonese, Korean, Indonesian, Japanese, Vietnamese, Russian). The utterances are recorded via different recording equipments and environments, making it a cross-channel LID task. Similarly to the official baseline model, we used the provided training data of the OLR  2020 Challenge (without the dialect data) for language enrollment. We omitted the datasets that were used as development test data in Task 1 or Task 2. In Task 1, the development test data was the Task 2 data of AP19-OLR (AP19-OLR-channel). This is consistent with the official baseline system developed for the OLR 2020 Challenge.

Task 3 of OLR 2020 Challenge is a closed-set identification task under noisy conditions. The language of each utterance is one of the five target languages (Cantonese, Korean, Japanese, Mandarin Chinese and Russian). The enrollment data for this task is all of the OLR 2020 Challenge training data for the particular five languages. 

As the third and a very different experimental dataset, we used the Lwazi corpus \cite{feld2009multilingual}. It contains audio samples from the 11 ofﬁcial languages in South Africa: Afrikaans, English, isiNdebele, isiXhosa, isiZulu, Sepedi, Sesotho, Setswana, Siswati, Tshivenda, and Xitsonga \cite{feld2009multilingual}. This data
was collected in South Africa over the telephone and is stored using 8 kHz sample rate. Many of the utterances in the corpus are very short which makes the data challenging for LID models.
The test and training data for the experiments was composed as follows: first, we selected 150 random speakers from the dataset, and included all the utterances from those speakers into the test set. The training set was composed by randomly selecting 2 hours of speech per language while excluding the utterances of the speakers in the test set. This training and test split is similar as was done in \cite{feng2019low}, except that we use a larger test set. For transfer learning experiments, the data was resampled to 16 kHz.

\subsection{What to use as a pretrained model?}

\begin{table*}[]
\caption{Comparison of different pretrained models on three datasets. In all cases, the resulting model is used without finetuning on the target training data.}
\label{tab:pretraining}
\begin{tabular}{l|rr|rr|rr}
\toprule
                                                                          & \multicolumn{2}{l|}{\textit{OLR 2020 Task 1}} & \multicolumn{2}{l|}{\textit{OLR 2020 Task 3}} & \multicolumn{2}{c}{\textit{Lwazi}}        \\
\midrule
\textit{Embeddings}        & \textit{$C_{\rm avg}$}              & \textit{EER}             & \textit{$C_{\rm avg}$}              & \textit{EER}             & \textit{$C_{\rm avg}$}   & EER   \\
\midrule
Official baseline (no pretraining) \cite{olr2020, feng2019low}                                       & 0.1321            & 14.58           & 0.0715            & 7.14            &        & 35.30 \\
Our baseline (Resnet + MHA, no pretraining)                                               & 0.1105            & 11.87           & 0.0709            & 6.80            & 0.2572 & 26.36 \\
Resnet, trained using VoxLingua107                                        & 0.0249            & 2.76            & 0.0147            & 1.33            & 0.1185 & 12.41 \\
WenetSpeech ASR Conformer enc., finetuned on VoxLingua107              & 0.0071            & 0.83            & 0.0043            & 0.39            & 0.0947 & 9.68  \\
CV multiling. ASR Conformer enc., finetuned on VoxLingua107 & 0.0061            & 0.59            & 0.0089            & 0.78            & 0.0985 & 10.13 \\
XLSR-53 wav2vec2.0, finetuned to VoxLingua107                                        & 0.0052            & 0.56            & 0.0057            & 0.57            & 0.1319 & 13.73 \\
XLS-R-300M  wav2vec2.0, finetuned on VoxLingua107                                     & \textbf{0.0021}            & \textbf{0.21}            & \textbf{0.0029}            & \textbf{0.28}            & \textbf{0.0809} & \textbf{8.30} \\
\bottomrule
\end{tabular}
\end{table*}

In this section, we compare different pretrained models as starting points for training a language embedding model on VoxLingua107. We experimented with the following models:
\begin{itemize} 
    \item Resnet model, randomly initialized (i.e., no pretraining before training with VoxLingua107)
    \item The encoder part of the Conformer-based sequence-to-sequence ASR model\footnote{\url{https://huggingface.co/espnet/pengcheng_guo_wenetspeech_asr_train_asr_raw_zh_char}} trained on WenetSpeech \cite{zhang2021wenetspeech}, a 10K+ hour corpus of Mandarin, containing speech from YouTube and podcasts. 
    \item The encoder part of the Conformer-based  sequence-to-sequence multilingual ASR model\footnote{\url{https://huggingface.co/espnet/Shinji_Watanabe_open_li52_asr_train_asr_raw_bpe7000_valid.acc.ave}} trained on the Mozilla CommonVoice corpus \cite{ardila2019common}, covering 52 languages.
    \item XLSR-53, the multilingual wav2vec2.0 model \cite{tjandra2021improved}.
    \item XLS-R-300M, an improved version of XLSR-53. It is trained on half a million hours of publicly available speech data in 128 languages, including the VoxLingua107 corpus.
\end{itemize}
Results on the three test datasets are listed in Table \ref{tab:pretraining}. In all cases, the resulting model is used for extracting embeddings for the language enrollment data and the final classifier is built using LDA and logistic regression.
We did not finetune the models on the target language enrollment data, except in the ``no pretraining'' case where only the enrollment data was used for training the embedding model. As we already experienced with the OLR 2020 Challenge Task 3 results, transfer learning from ASR encoder or wav2vec2.0 models results in large gains, compared to the Resnet model trained on VoxLingua107 from scratch. It is surprising to see that a monolingual Conformer model trained on WenetSpeech results in comparable results to the multilingual ASR model and XLSR-53. This suggests that having diverse real-world data among the pretraining data is as important as multilinguality, as WenetSpeech contains speech ``from the wild'' while CommonVoice has only prompted speech. Finally, we see that the new multilingual wav2vec2.0 model gives the best results across all datasets.  We also see that pretraining is much more effective on the OLR data than on the Lwazi corpus. This can be probably explained by two factors: first, only Afrikaans and English among the 11 target languages in Lwazi are covered by VoxLingua107. Second, all the pretrained models are trained on (mostly) 16 kHz data, making them less effective for 8 kHz telephone speech.

\subsection{VoxLingua107: does size or variety matter?}

\begin{figure}[tb]
  \centering
  \includegraphics[width=0.90\linewidth]{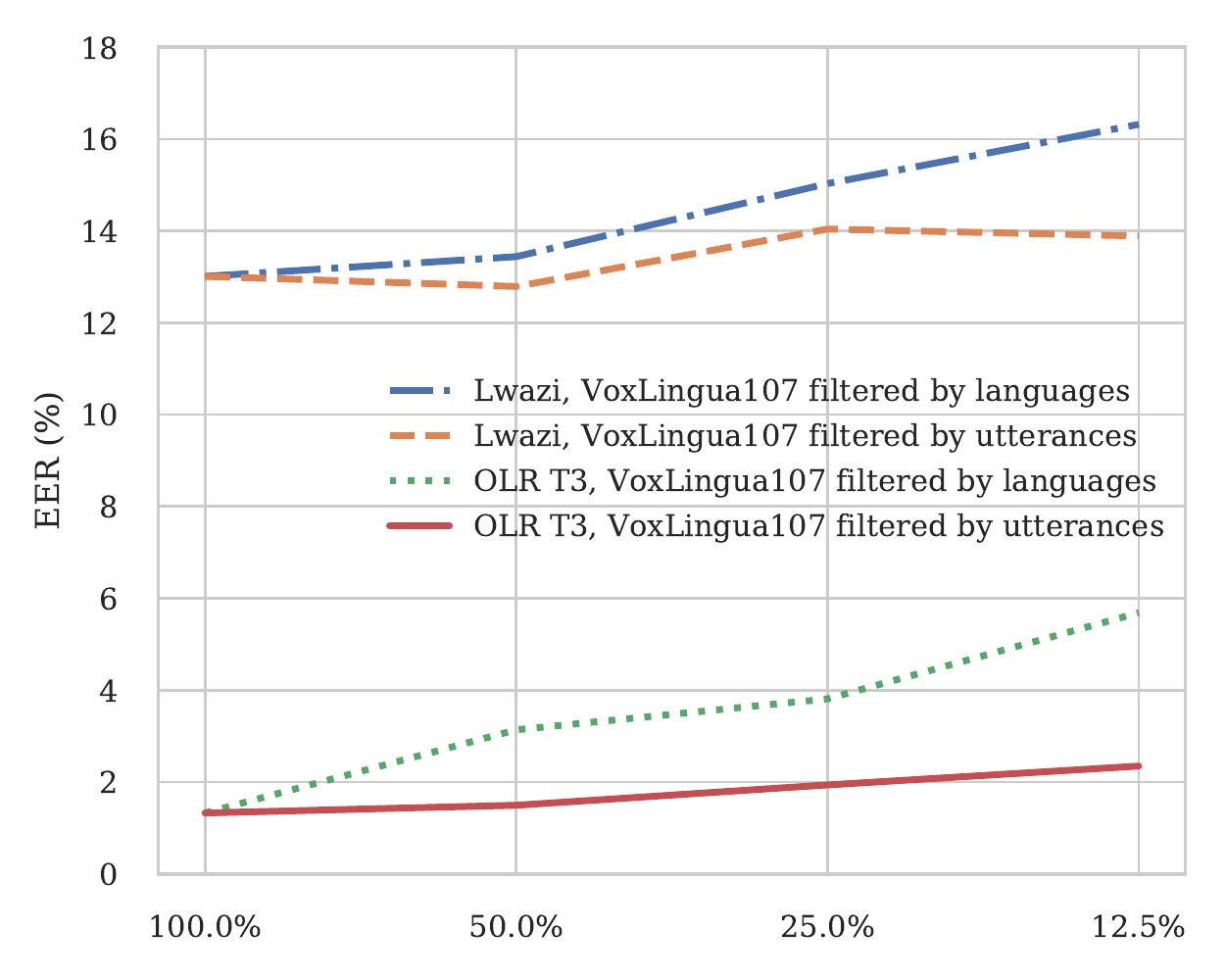}
  \caption{Language identification performance when the amount of pretraining data is reduced, by either removing random utterances (but keeping all the languages) or by removing whole languages.}
  \label{fig:ablation}
\end{figure}

Using the full VoxLingua107 dataset for pretraining resulted in substantial improvements throughout the experiments. Here we analyze the effect of the size and language variety of the pretraining data on transfer learning performance. This is done by training Resnet-based language embedding models on various subsets of VoxLingua107. In order to factor out external effects, we do not use pretrained Conformer and wav2vec2.0 models here.

We created several subsets of the VoxLingua107 data, each containing either 50\%, 25\% or 12.5\% of the original data. The subsetting was performed in two different ways: either per utterance or per language. In the first case, a certain proportion of utterances were kept while keeping the number of languages intact, while in the second case, all utterances of randomly selected languages were kept, so that the number of utterances would be roughly  the same as in the first case.

Figure \ref{fig:ablation} shows the EER on OLR 2020 Task 3 and Lwazi data, when using the described subsets of VoxLingua107 for pretraining. No finetuning was performed in those experiments. It is clear from the experiments that the number of different languages present in the pretraining data is more important than the number of utterances.

In the case of OLR Task 3 experiments, the fast performance drop when pretraining languages are removed can be partly explained by the fact that the language subsets drop some of the OLR Task 3 target languages (Cantonese, Korean, Japanese, Mandarin Chinese and Russian). Indeed, the 50\% language subset includes only Korean, and the even smaller subsets contain none of the target languages. Therefore, we trained another model on only the four target languages (except Cantonese) of VoxLingua107 and used this as the embedding extractor. This resulted in EER of 8.1\% which is worse than the model trained on the 14 language (12.5\%) subset of VoxLingua107 where none of the target languages is present (EER of 5.7\%). This suggests that although having the target languages among the pretraining languages is beneficial, it's more important to pretrain on a large variety of languages.

\subsection{How much data do we need for training the backend?}

\begin{figure}[tb]
  \centering
  \includegraphics[width=\linewidth]{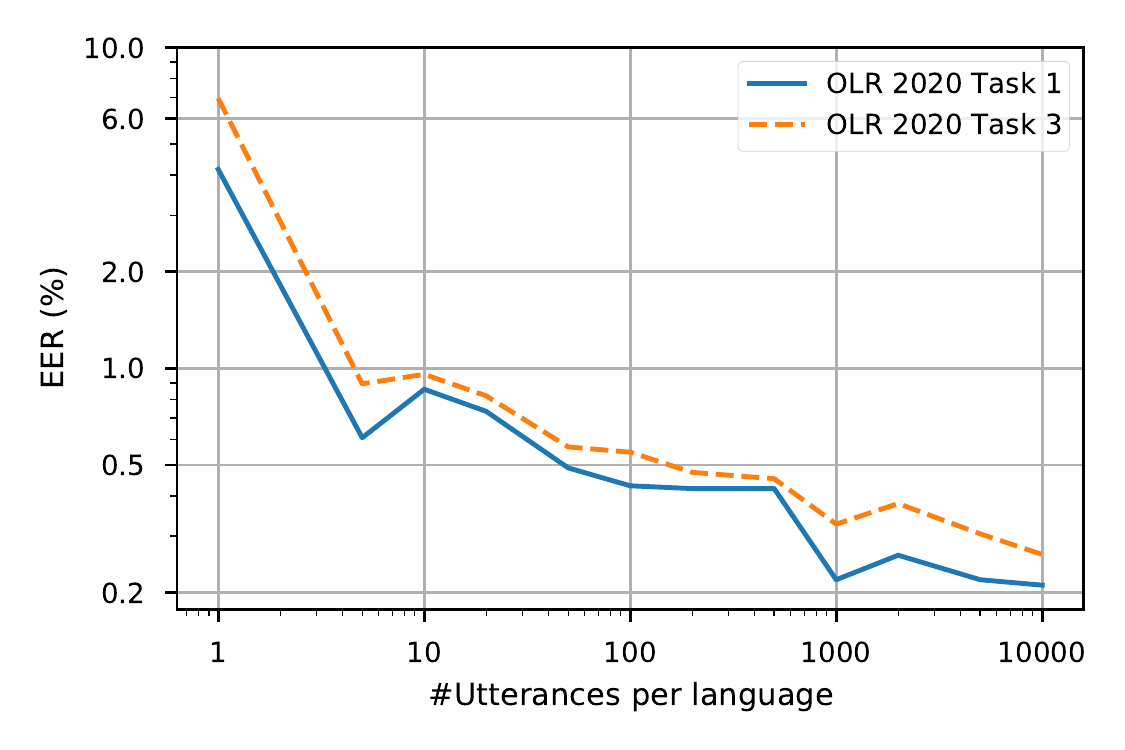}
  \caption{Number of enrollment utterances per language \textit{vs} EER. Note the log axes.}
  \label{fig:utts_vs_eer}
\end{figure}

We investigated how much enrollment data is needed for training the logistic regression based backend model with the best embedding model based on XLS-R-300M and VoxLingua107. For this, we created random subsets of OLR 2020 Task 1 and Task 3 training data with increasing number of utterances per target language. Figure \ref{fig:utts_vs_eer} shows that with just one utterance per language, this approach already outperforms the baseline model that doesn't use pretraining, but uses around 10\,000 training utterances per language. With only five utterances per language, we already achieve EER rates of below 1\%. 

\section{Conclusion}

We described the TalTech spoken language identification systems for the OLR 2021 Challenge.  For the constrained task, our best-performing single system was trained via transfer learning from the encoder part of a Conformer-Transformer based multilingual model. For the unconstrained task, we relied on two important external resources: the XLSR-53 pretrained multilingual wav2vec2.0 model, and the VoxLingua107 corpus. Our best performing model was trained via transfer learning from XLSR-53 using two finetuning steps: first using VoxLingua107, followed by target language training data.

Post-evaluation experiments showed that both ASR encoder models as well as self-supervised wav2vec2.0 models serve as good starting points for training a language embedding extractor on the VoxLingua107 dataset. 
We also showed that the variety of languages in VoxLingua107 is more important than the total amount speech. Finally, we demonstrated that on cross-channel and noisy test sets of OLR 2020, using pretrained embeddings requires only five language enrollment utterances to achieve below 1\% EER.

%

\bibliographystyle{IEEEbib}

\bibliography{mybib}

\end{document}